\documentstyle[11pt,newpasp,twoside,epsf]{article}
\def\edcomment#1{\iffalse\marginpar{\raggedright\sl#1\/}\else\relax\fi}
\reversemarginpar

\begin{document}
\title{The Detection of New Methanol Masers in the $\bf 5_{-1}-4_0E$ Line}
\author{S.V. Kalenskii, V.I. Slysh, I.E. Val'tts}
\affil{Astro Space Center, Lebedev's Physical Institute,
Profsoyuznaya str. 84/32, 117810 Moscow, Russia}
\author{A. Winnberg and L.E.B. Johansson}
\affil{Onsala Space Observatory, S-439 92 Onsala, Sweden}

\begin{abstract}
 Fifty-one object in the $5_{-1}-4_0E$ methanol line at 84.5~GHz was detected
during a survey of Class I maser sources. Narrow maser features were found
in 17 of these. Broad quasi-thermal lines were detected towards other sources.
One of the objects with narrow features, the young bipolar outflow
L~1157 was also observed in the $8_0-7_1A^+$ line at 95.2~GHz;
a narrow line was detected at this frequency. Analysis showed that the broad
lines are usually inverted. The quasi-thermal profiles imply that
the line opacities are not larger than several units. These results confirm
the plausibility of models in which compact Class I masers appear in extended
sources as a result of an appropriate velocity field.

Measurements of linear polarization at 84.5~GHz in 13 sources
were made. No polarization was found except a tentative detection
of a weak polarization in M~8E.
\end{abstract}

\section{Introduction}
 Methanol masers
in the $5_{-1}-4_0E$ line at 84521.21~MHz were found by Batrla and
Menten~(1988) and Menten~(1991) towards NGC~2264, OMC-2, and DR~21,
but no extended survey in this line had been done. The $5_{-1}-4_0E$
transition belongs to the Class~I (Menten, 1991). Its excitation is similar
to that of the $4_{-1}-3_0E$ and $6_{-1}-5_0E$ transitions.
Since methanol masers emit in several lines of the same
class, we expect the detection of a fairly large number of maser sources
at 84.5~GHz. Their parameters should be taken into account when modeling
maser sources. Therefore, we made a survey of known Class~I maser sources
at 84.5~GHz.

\section{Observations and results}

The observations were carried out in May 1997 and March 2000
with the \mbox{20-m} millimetre-wave telescope of the Onsala Space Observatory.
A sample of 13 sources at 84.5~GHz was observed in June 2000 with the 12-m
NRAO telescope at Kitt-Peak in remote mode from Astro Space Center.

\begin{figure*}
\plotfiddle{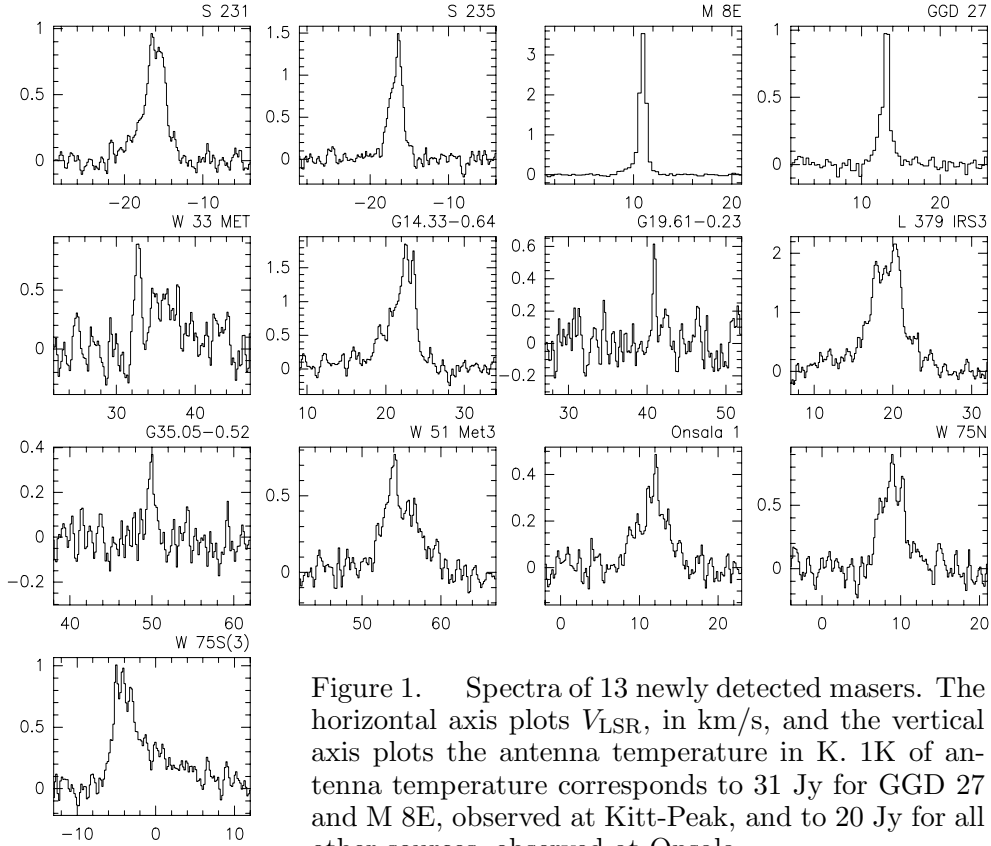}{90mm}{0}{70}{75}{-20}{-290}
\hspace{28mm}
\parbox{105mm}{\caption{Spectra of 13 newly detected masers. The horizontal axis plots
$V_{\rm LSR}$, in km/s, and the vertical axis plots the antenna
temperature in K. 1K of antenna temperature corresponds
to 31~Jy for GGD~27 and M~8E, observed at Kitt-Peak, and
to 20~Jy for all other sources, observed at Onsala.}}
\end{figure*}

Emission was detected in 51 of the 54 sources observed.
The spectra are markedly different from those of the strongest Class~I
transition, $7_0-6_1A^+$ at 44.1~GHz. At 44.1~GHz, most of the sources from
our sample have bright and narrow maser features, whereas broad
quasi-thermal components dominate at 84.5~GHz, and narrow ($< 1.5$ km/s)
features are present in the spectra of only 17 of the 51 detected sources
(Fig. 1). However, it is possible that at least some of the quasi-thermal
lines contain narrow maser components.

The shape of the 84.5~GHz spectra closely resembles the shape of the spectra
of the same sources in the $8_0-7_1A^+$ (Val'tts~et~al.~1995)
and $6_{-1}-5_0E$ (Slysh~et~al.~1999) transitions at 95.2 and 132.8~GHz,
respectively. 
The relationships between
the integrated intensities of thermal lines at 84.5, 95.2 and 132.8~GHz
can be fitted by the equations
\begin{equation}
\int T_{mb}dV (95.2) = 0.4\int T_{mb}dV (84.5) + 0.17
\label{cor1}
\end{equation}
and 
\begin{equation}
\int T_{mb}dV (132.8) = 0.7\int T_{mb}dV (84.5) + 0.0
\label{cor2}
\end{equation}
Here $T_{mb}$ is the main-beam brightness temperature. The relative decrease
of the line intensities at 132.8, and especially at 95.2~GHz, is probably
connected with the decrease of level population with increase of their
energies: at a gas temperature of 35~K the population of the $8_0A^+$ level
is about 40\% of the population of the $5_{-1}E$ level, making it
possible to explain the relationships obtained.

Note the detection of narrow features at 84.5 and 95.2~GHz
towards the young bipolar outflow L~1157.
Unlike other methanol masers, which are associated with
high-luminosity young stellar objects (above $10^3\; L_{\sun}$),
this one is associated with an object of low luminocity
($11\; L_{\sun}$).

\section{Excitation temperature of the quasi-thermal lines} 

Slysh et al. (1999) showed that even quasi-thermal $6_{-1}-5_0E$ lines are
typically inverted and their quasi-thermal appearance indicates that the line
opacities are not large enough to cause significant narrowing. Since
the excitation of the $5_{-1}-4_0E$ transition is similar to that of the
$6_{-1}-5_0E$ transition it is possible that the quasi-thermal $5_{-1}-4_0E$
lines are also inverted. To test this hypothesis, we determined
the excitation temperature of the $5_{-1}-4_0E$ lines using the intensities
of the $4_0-4_{-1}E$ lines at 157.2~GHz, measured by Slysh~et~al.~(1999).
The excitation temperatures were derived analytically using a simple
method described by Slysh et al.~(1999).
We applied this method to 20 quasi-thermal sources, and
for each, obtained negative excitation temperature between
$\approx -1.5$~K and $\approx -4.5$~K, i.e., the $5_{-1}-4_0E$
quasi-thermal lines proved to be strongly inverted.

The excitation temperatures derived in this way are distorted by a number
of factors, such as the line opacities, influence of microwave background etc
(Slysh~et~al.,~1999). Therefore, we verified the results using a grid of LVG
methanol models spanning the ranges $10^4-10^8$~cm$^{-3}$ in density, 
10--100~K in temperature, and 
$7\times 10^{-7}-2\times 10^{-3}$ cm$^{-3}$/(km/s pc$^{-1}$)
in methanol density divided by the velocity gradient. For each source,
we selected the models corresponding to the observed ratios of the
main-beam brightness temperatures of the $5_{-1}-4_0E$ line and
the $4_0-4_{-1}E$ and $5_0-5_{-1}E$ lines, observed by Slysh~et~al.~(1999).
The results are as follows:

For the majority of the sources, we found that only models with inversion of
the $5_{-1}-4_0E$ transition or models with unlikely high
methanol abundances satisfy the observed line ratios. 
In G29.95-0.02, G34.26+0.15, NGC~7538, W~49N, and W~51E1/E2, the observed
intensity ratios can be obtained both in models with the inversion and in
realistic models with positive excitation temperatures. 
However, since a number of models with inversion (i.e., same as those
for the other 15 sources) are applicable to these objects as well, it is
not clear whether they are somehow different from the others or not.

Thus, the quasi-thermal $5_{-1}-4_0E$ methanol lines, like the $6_{-1}-5_0E$ 
lines, are typically inverted.
This result confirms
the plausibility of models in which compact Class I masers appear in extended
sources as a result of an appropriate velocity field
(see, e.g., Sobolev~et~al.~(1998).

\section{Search for linear polarization}
In the series of observations, performed in June 2000 with the 12-m NRAO
telescope at Kitt-Peak we tried to find linear polarization at 84.5~GHz
towards 13 sources. 
We expected that Class I methanol masers may arise in a gas permeated
by magnetic field and may exhibit a weak linear polarization similar to that
of some H$_2$O masers.

Two polarization channels of the 3-mm receiver at Kitt-Peak can
measure both senses of linear polarization simultaneously.
Different brightness temperatures, measured in different
channels would mean that the radiation is linearly polarized. One can test
whether the difference is a result of linear polarization by
tracing the source during a sufficiently long time range.
Owing to the diurnal rotation of the sky the direction of
the polarization plane 
will vary, resulting in regular variations of the measured
brightness temperatures in the two channels and hence, in a regular
variation of the difference between them.
                             
We failed to find any difference between channels in 12 sources and obtained
the upper limits for the degree of linear polarization
within the range 3\%--30\%. The only exception is a strong maser
M~8E. Here we found a small difference
between polarization channels, which might appear due to
a weak ($\approx~3\%$) linear polarization at 84.5~GHz.
Unfortunately, this southern source ($\delta=-24^\circ$) cannot be traced
along a significant part of its diurnal trajectory
from the northern hemisphere. Therefore, we could not detect
any regular variation of the difference between channels and cannot
state that the difference is a result of linear
polarization. Further polarization measurements of this source at 84.5~GHz
are required.

\acknowledgements
The authors are grateful to the staff of the Onsala Space
Observatory and to the staff of the Kitt Peak telescope for providing help
during the observations. The work was done
under a partial financial support of the Russian Foundation for Basic
Research (grant No. 95-02-05826) and INTAS (grant No. 97-1451). 

The resuts of the work are partly presented in Kalenskii et al.~(2001).


\begin{references}
\reference Batrla, W. \& Menten, K.M., 1988, \apj, 329, L117
\reference Kalenskii, S.V., Slysh, V.I., Val'tts, I.E., Winnberg~A.,
      \& Johansson L.E.B., 2001, Astronomy Reports, 45, 26
\reference Menten K.M., in: Publ. Astron. Soc. Pac., Skylines, Proc.
      3rd Haystack Observatory Meeting, eds. A.D.~Haschick, P.T.P.~Ho, 119
\reference Slysh V.I., Kalenskii S.V., Val'tts I.E., Golubev V.V.,
      Mead K., 1999, \apjs, 123, 515
\reference Sobolev, A.M., Wallin, B.K., \& Watson, W.D., 1998, \apj, 498, 763	  
\reference Val'tts, I.E., Dzura, A.M., Kalenskii, S.V., Slysh, V.I.,
      Booth R., \& Winnberg, A., 1995, Astronomy Reports, 39, 18 
\end{references}
\end{document}